 \documentclass[11pt,twocolumn]{article}

\newcommand{\half}{{\frac{1}{2}}}

\begin{document}

\begin{flushright}  TAUP-2749-03  \end{flushright}
\begin{center}
{\Large \bf Non-supersymmetric deformation of the Klebanov-Strassler
  model and the related plane wave theory}\\[12pt]
{Stanislav Kuperstein}\\[5pt]
{\sl  School of Physics and Astronomy\\
  The Raymond and Beverly Sackler Faculty of Exact Sciences\\
  Tel Aviv University, Ramat Aviv, 69978, Israel \\
E-mail: kupers@post.tau.ac.il}
\end{center}

%\begin{quote}
 {\small  We find a regular analytic 1st order deformation of the Klebanov-Strassler background.
 From the dual gauge theory point of view  the deformation describes 
supersymmetry soft breaking gaugino mass terms.
We calculate the difference in vacuum energies between the supersymmetric
and the non-supersymmetric solutions and find that it  matches the  field theory prediction.
We also discuss  the breaking of the 
$U(1)_R$ symmetry and the space-time dependence of the gaugino bilinears two point function.
Finally, we determine the Penrose limit of the non-supersymmetric
background and write down the corresponding plane wave string
theory. This  string describes ``annulons''-heavy hadrons with mass
proportional to large global charge. Surprisingly the string spectrum 
  has two fermionic zero modes.  This  implies that the   sector in
  the  non-supersymmetric gauge theory which   is the dual of the
  annulons is supersymmetric.} \\[24pt]

\section{Introduction}

Since the formulation of
the AdS/CFT conjecture \cite{Maldacena:1998re},
\cite{Gubser:1998bc}, \cite{Witten:1998qj} (see
 \cite{Aharony:1999ti} for a review) there has been great progress in the study of
theories with less
supersymmetries and not necessarily conformal.
There are several approaches one can use to break the $\mathcal{N}=4$
supersymmetry down to $\mathcal{N}=2$ or $\mathcal{N}=1$.

A few years ago two important examples of supergravity duals of $\mathcal{N}=1$  
gauge theories have been  provided by  \cite{Maldacena:2000yy}
and  \cite{Klebanov:2000hb}
(see \cite{Bertolini:2003iv} and \cite{Bigazzi:2003ui} for recent reviews). 
The Maldacena-Nunez (MN) background
consists of NS5-branes wrapped on an $S^2$ and based on the solution
of  \cite{Chamseddine:1997nm}. 
The supergravity dual of Klebanov-Strassler (KS) involves D5 branes wrapped around
a shrinking $S^2$. The metric has a standard D3-form with the
6d deformed conifold being the transversal part of the 10d space.

Non-supersymmetric deformations of the MN background have been studied by number of authors.
In  \cite{Aharony:2002vp} the supersymmetry was broken completely by giving masses for some of the
scalar fields. It was argued that the deformed non-supersymmetric 
background is guaranteed to be stable, since the original dual gauge theory had a
mass gap. On other hand, the authors of  \cite{Evans:2002mc} used the
solution of \cite{Gubser:2001eg} to study
the supersymmetry breaking by the inclusion of a gaugino mass term and a condensate.  
Evidently, the global symmetry remains unbroken under this deformation.

Our main goal is to find a non-singular,
non-supersymmetric deformation of the KS solution,
which preserves the global symmetries of the original background and to study the Penrose limit
of the new solution. 
The problem has
been already attacked by different authors \cite{Apreda:2003gc},\cite{Borokhov:2002fm}.
The authors of \cite{Borokhov:2002fm} suggested a computational technique for studying the
non-supersymmetric solution. The technique is based on the
modification of the first order BPS equations, so that we might continue to use 
a superpotential even for a non-supersymmetric solution. 
In short, one obtains
a set of sixteen 1st equations and one zero-order constraint
instead of eight standard 2nd order differential  equations.

In this paper (see \cite{Kuperstein:2003yt} for more comprehensive discussion)
we determine and describe  a regular \emph{analytic} solution of the 1st order equations 
similar to those appearing in \cite{Borokhov:2002fm}. 
We note that these equations are significantly 
simplified once we properly redefine the radial coordinate.
(The equations transform non-trivially under the coordinate redefinition since
one has to apply the ``zero-energy'' constraint, which removes the ``gauge freedom'' of the
coordinate transformation).
We also demonstrate how part of the 1st order equations can be re-derived using 
the usual 2nd order IIB equations of motion.

Our solution preserves the global symmetry and therefore
describes a deformation corresponding to the inclusion of mass terms
of the two gaugino bilinears in the dual gauge theory.

We construct a Penrose limit (see \cite{Gueven:2000ru}, \cite{Blau:2001ne},
\cite{Metsaev:2001bj}, \cite{Berenstein:2002jq},
\cite{Metsaev:2002re} and \cite{Blau:2002mw})
of our non-super\-sym\-met\-ric 
KS background and obtain a pp-wave metric
and a complex 3-form which are very  similar to the PL limit \cite{Gimon:2002nr}
of the supersymmetric solution.

We also quantize the light-cone string Hamiltonian and determine the
masses  of the bosonic and fermionic modes. These masses, though
different from the supersymmetric case, still obey the relation that
the sum of the mass squared is the same for bosonic and fermionic
modes.
Again the  string describes kinematics and excitations of heavy
hadrons (called ``annulons'' \cite{Gimon:2002nr}) with masses proportional to a large global 
symmetry charge $M_{annulon}=m_0 J$. The only difference between them
and those of \cite{Gimon:2002nr} is a modification of $m_0$.
A  surprising feature of the string spectrum is that, like in the
Penrose limit of the KS background, here as well, there are 
two fermionic zero modes.
In the dual field theory this implies that even though the full theory is non-supersymmetric,
the sector of states with large $J$ charge  admits supersymmetry.
It is conceivable that  in this limit of states of large mass the 
impact of the small  gaugino mass deformations is washed away.

The authors of \cite{Apreda:2003gs}
used the  solution of \cite{Apreda:2003gc} to take the PL.
The IR expansion of the fields given in \cite{Apreda:2003gc} differs, however,
from our solution (see later) and therefore the pp-wave background
of \cite{Apreda:2003gs} is also slightly different from the metric we have
derived.

\section{The Klebanov-Strassler model and beyond}
\label{section:TKSmab}

Before reviewing the main features of the KS solution it will be worth to
write down  the type IIB equation of motion for a case of  
a constant dilaton ($e^\Phi =g_s$), a vanishing
axion ($C_0=0$) and with the 10d metric
and the 5-form flux having the structure of the D3-brane solution, namely:
\begin{equation}              \label{eq:metric}
ds^2 = h^{-1/2} \mathbf{d} x_\mu^2 +
       h^{1/2} \d s^2_{M_6}
\end{equation}
and
\begin{equation}              \label{eq:5form}
\tilde{F}_5 = \frac{1}{g_s} (1 + \star_{10}) \mathbf{d} h^{-1} \wedge \mathbf{d} x_0 \wedge \ldots
                               \wedge \mathbf{d} x_3,
\end{equation}
where $M_6$ is a 6d Ricci flat transversal space and the harmonic
function $h$ depends only on the coordinates on $M_6$.
We will denote the Hodge dual on  $\mathnormal{M}_6$ by $\star_6$.
In order to find the connection between the 3-forms and the warp
function $h$ we have to use the 5-form equation. We end up with:
\begin{eqnarray} \label{eq:dsdh}
\tilde{F}_5 =  B_2 \wedge F_3 + \star_{10} \left( B_2 \wedge  F_3\right)
\nonumber \\
\quad \textrm{or} \quad
 \mathbf{d} h = - g_s \star_6 \left( B_2 \wedge F_3 \right),
\end{eqnarray}
where the 2nd equation is the integrated version of the first.
Next we consider the 3-forms equations. Applying (\ref{eq:dsdh}) and the relation between 
$F_5$ and $\tilde{F}_5$ we get:
\begin{equation}           \label{eq:F3H3}
\mathbf{d} \left[ h^{-1} \left( \star_6  F_3  +  \frac{1}{g_s}  H_3  \right) \right] =0
\end{equation}
and similarly for $H_3$.
In deriving this result we have used the fact that
all the forms have their legs along the 6d space.
Finally, 
calculating the Ricci scalar of the metric (\ref{eq:metric}) we
re-write the metric equation of motion:
\begin{equation}        \label{eq:Rmn}
\mathbf{d} \star_6 \mathbf{d}  h  = \frac{1}{2} \left[ H_3 \wedge \star_6 H_3 
                                +  g_s^2 F_3 \wedge \star_6 F_3 \right]. 
\end{equation}
The equations we have written
(\ref{eq:dsdh},\ref{eq:F3H3},\ref{eq:Rmn})
as well as the the dilaton and the axion equations  
are easily solved by
requiring that:
\begin{eqnarray}        \label{eq:dualFH}
 \star_6  F_3  =  - g_s^{-1}  H_3
\quad
\textrm{and}
\quad
 \star_6  H_3  =  g_s  F_3.
\end{eqnarray}
In this case the complex form $G_3 \equiv F_3 + \frac{i}{g_s} H_3$ is
imaginary self dual $ \star_6 G_3 =i G_3$.

Note that
the equation for $h$ is a first order differential equation, even
though the solution is not supersymmetric in general.

The most important example of the supersymmetric solution is the
Klebanov-Strassler model \cite{Klebanov:2000hb}, where the 6d manifold is the deformed
conifold space.
The $\mathbf{M}$ fractional D5-branes wrapping the shrinking $S^2$ are introduced through the RR 3-form
and on using the duality relations (\ref{eq:dualFH}) one may also find the NS 3-form:
\begin{eqnarray}     \label{eq:F3andH3}
H_{3} &=& g_s \mathbf{M} \mathbf{d} \left[ 
                     f(\tau)   g^1 \wedge g^2 +
                     k(\tau)   g^3 \wedge g^4 
          \right],
\nonumber \\
F_{3} &=& \mathbf{M} \Big[  g^5 \wedge g^3 \wedge g^4 + 
\nonumber \\
&&  + \mathbf{d} \left(  F(\tau) g^1 \wedge g^3 + g^2 \wedge g^4 \right)
          \Big]
\end{eqnarray}
where $\tau$ is the radial coordinate and the functions
$f(\tau)$,$k(\tau)$ and $F(\tau)$ satisfy a set of three first order
differential equations \cite{Klebanov:2000hb}.
This set has three dimensional space of solutions.
Using the complex structure of 
the deformed conifold space \cite{Papadopoulos:2000gj}
the complex form $G_{3}= F_{3}+ \frac{i}{g_s} H_{3}$
can be identified for the Klebanov-Strassler solution as a regular $(2,1)$ form. 
There are also two additional solutions corresponding to a $(0,3)$ form 
which breaks the supersymmetry and diverges at $\tau \to \infty$
and a $(2,1)$ form which is singular at $\tau=0$.

The dual field theory realized on the world-volume of the $\mathbf{N}$ physical and $\mathbf{M}$ 
fractional D3-branes is a 4d $\mathcal{N}=1$ supersymmetric 
$SU(\mathbf{N}+\mathbf{M}) \times SU(\mathbf{N})$ gauge theory with a $SU(2) \times SU(2)$
global symmetry inherited from the conifold isometries.
The gauge theory is coupled to two bi-fundamental
chiral multiplets $A$ and $B$, which transform as a doublet of one of
the $SU(2)$'s each and are inert under the second $SU(2)$.
This theory is believed to exhibit a cascade of Seiberg dualities
reducing in the deep IR to pure $SU(\mathbf{M})$.
On the supergravity side $\mathbf{M}$ is fixed by the charge of the RR 3-form,
while $\mathbf{N}$ is encoded in the UV behavior of the 5-form.
The sum of the gauging couplings is constant and the logarithmic running of
the difference is determined by the NS 2-form.

Similarly to pure $SU(\mathbf{M})$ the theory confines. This is evident by virtue of the fact that the warp
factor approaches a constant value $h_0 \sim a_0$ at $\tau \to 0$ and therefore the tension of the confining strings
does not diverge. This conclusion is valid only for a non-zero value of the deformation parameter $\epsilon$,
since $a_0 \sim \epsilon^{-8/3}$. 
Note also that for $\epsilon \neq 0$ the $U(1)_R$ conifold symmetry is
broken down to $\mathbf{Z}_2$. This is the symmetry preserved by the gaugino bilinear
$\textrm{Tr} \lambda\lambda(x)$.
In the supergravity dual this gauge theory operator is associated with the form 
$C_2 = C^{RR}_2 + i B^{NS}_2$ \cite{Loewy:2001pq}. Subtracting the asymptotic value of $G_3=\mathbf{d} C_2$
we find at $\tau \to \infty$:
\begin{eqnarray}     \label{eq:DG}
&\Delta G_3 \approx \frac{\mathbf{M}}{2} \tau e^{-\tau} \omega_3, \,    
\Delta C_2 \approx - \frac{\mathbf{M}}{2} \tau e^{-\tau} \omega_2, &
\nonumber \\
&\omega_3 = \mathbf{d} \omega_2, \, \omega_2 = \Big[ \left( g^1 \wedge g^3 + g^2 \wedge g^4 \right) +&
\nonumber \\
&+ i g_s  \left( g^1 \wedge g^2 - g^3 \wedge g^4 \right)  \Big], &      
\end{eqnarray}
where we write only the polarization along $T_{1,1}$
and we see that $\Delta C_2$ transforms under $U(1)_R$ by the same phase as 
$\textrm{Tr} (\lambda\lambda)$. Moreover, $\Delta G_3$ has an asymptotic behavior 
we would expect from a scalar operator of dimension 3 and a non-zero VEV, namely:
$\Delta G_3 = \half \mathbf{M} \frac{m^3}{r^3} \ln \frac{r^3}{m^3} \omega_3$,
where the deformation parameter is related to the 4d mass scale through $m \sim \epsilon^{2/3}$.

Finally, we will recall the identification of supergravity fields with   
gauge theory operators. In order to find this correspondence one writes the most
general $SU(2) \times SU(2)$ invariant background ansatz, which includes the supersymmetric
KS solution:
\begin{eqnarray}          \label{eq:ansatz}
& ds^2 = 2^{1/2} 3^{3/4} \Big[  e^{-5q+2Y} ( d x_{\mu})^2+   &
\nonumber \\
&   + \frac{e^{3q-8p}}{9} \left( d\tau^2 + g_5^2 \right) 
      +  \frac{e^{3q+2p+y}}{6} \left( g_1^2 + g_2^2 \right) + & 
\nonumber\\         
& +   \frac{e^{3q+2p-y}}{6} \left( g_3^2 + g_4^2 \right)\Big] , \quad \Phi= \Phi(\tau),&
\end{eqnarray}
with the 3-forms are given by (\ref{eq:F3andH3}) and the 5-form by (\ref{eq:5form}).
This general ansatz includes both the conformal solution with a singular geometry
 ($y=\tilde{f}-\tilde{k}=0$) and the non-conformal case with regular deformed conifold 
($y,\tilde{f}-\tilde{k} \neq 0$). Here $\tilde{f}, \tilde{k}$ and $\tilde{F}$  
are the rescaled KS functions: 
$ \tilde{f} = - 2 P g_s f, \,  \tilde{k} = - 2 P g_s k,  \,
\tilde{F} =  2 P F $
and the constant $P$ is related to the number of fractional branes:
$P= \frac{1}{4} \mathbf{M}l_s^2$.
Note that for the given structure of the 3-form $F_3$ the integral $\int_{S_{3}} F_3$
does not depend on $\tilde{F}(\tau)$. Moreover, the NS-NS 3-form has the same structure
as in the KS solution as dictated by the equation for a vanishing axion $H_3 \wedge F_3=0$.

Integration of the type IIB Lagrangian over the angular and the world-volume coordinates 
yields a 1d effective action:
\begin{eqnarray}
S \sim \int d \tau \left( - \half G_{ij} \dot{\phi}^i \dot{\phi}^j - V(\phi) \right),     
\end{eqnarray}
and we refer the reader to
\cite{Klebanov:2000nc},\cite{PandoZayas:2000sq},\cite{Papadopoulos:2000gj}, \cite{Bigazzi:2000uu}
for an explicit form of the metric and the potential $V(\phi)$.
There is also a ``zero-energy'' constraint
$ \half G_{ij} \dot{\phi}^i \dot{\phi}^j - V(\phi) = 0$.
This Lagrangian admits a superpotential
\begin{eqnarray}
&V = \frac{1}{8}  G^{ij}\partial_i W \partial_j  W 
\quad \textrm{for}& 
\nonumber \\
&   W= - 3 e^{4Y+4p-4q} \cosh y - &
\nonumber \\
&- 2 e^{4Y-6p-4q} - 3 \sqrt{3} e^{4Y-10q} \tilde{L} &
\end{eqnarray}
and for supersymmetric solutions the second order equations of motion
can be reduced to the first order ones:
\begin{equation}
\dot{\phi^i} = \half G^{ij} \partial_j W.
\end{equation}
 The potential appearing in the action 
has an $\mathcal{N}=1$ critical point corresponding to the conformal background
$AdS_5 \times T_{1,1}$ generated by physical D3-branes in absence of fractional branes ($P=0$).
Expanding the potential around the critical point and using the mass/ dimension formula
$\Delta = 2 + \sqrt{4 +m^2}$ one obtains the dimensions of the fields,
which now can be identified with  various gauge theory operators
\cite{Ceresole:1999zs}, \cite{Bigazzi:2000uu}. 
Here we list two of them (both with $\Delta=3$):
\begin{eqnarray}
&\xi_2 \sim -F+\frac{k-f}{2} \to \textrm{Tr} \left( W_{(1)}^2+W_{(2)}^2 \right),&
\nonumber \\
&y  \to \textrm{Tr} \left( W_{(1)}^2 - W_{(2)}^2 \right)&
\nonumber
\end{eqnarray}
There are also two massless fields. $s=f+k$ is associated with a marginal direction in the CFT 
and the corresponding operator is $\textrm{Tr}\left(F_{(1)}^2-F_{(2)}^2\right)$. Similarly, 
the dilaton $\Phi$ corresponds to 
$\textrm{Tr}\left(F_{(1)}^2+F_{(2)}^2\right)$.

In this paper we will focus on the non-supersymmetric deformation
of the KS background by introducing mass terms of the gaugino bilinears 
associated with both $\xi_2$ and $y$. The former field is related to the 
SUGRA 3-forms and the latter is responsible for a deformation of the 6d metric.
The expected UV behavior of the fields in the background deformed by the masses is
$g(\tau) e^{-\tau/3}$, where $g(\tau)$ is a polynomial in $\tau$.

\section{Non-supersymmetric extension of KS}
\label{section:NseoKS}

We start this section with a brief review of the method proposed by
\cite{Borokhov:2002fm}
(see also \cite{Ferrara:1995ih}, \cite{Freedman:1999gp},
\cite{Skenderis:1999mm}, \cite{DeWolfe:1999cp}, \cite{deBoer:1999xf} and \cite{Wijnholt:1999vk})
to study  first order non-supersymmetric deformations of the KS background
still making use of the superpotential.
We expand the fields around a given supersymmetric solution derived from the superpotential
$\phi^i = \phi^i_0 + \delta \cdot \bar{\phi}^i + O(\delta^2)$.
Define new functions:
\begin{eqnarray}
& \xi_i = G_{ij}(\phi_0) \left( \frac{d \bar{\phi}^j}{d \tau} - M^j_k(\phi_0) \bar{\phi}^k \right)
\quad
\textrm{where} &
\nonumber \\
& M^j_k = \half \frac{\partial}{\partial \phi^k} \left( G^{jl} \frac{\partial W}{d \phi^l} \right) .& 
\end{eqnarray}
Now one might represent the linearized equations of motion as a ``double'' set 
of first order equations (we refer the reader to \cite{Borokhov:2002fm} for the proof):
\begin{eqnarray}   \label{eq:1st}
\frac{d \xi_i}{d \tau} + \xi_j M^j_i= 0,
\quad
\frac{d \bar{\phi}^i}{d \tau} - M^i_j \bar{\phi}^j = G^{ik} \xi_k
\end{eqnarray}
while the zero-energy condition can be rephrased as
$\xi_k \partial_\tau \bar{\phi}^k  = 0$.

An important remark is in order. 
One can use various definitions 
for the radial coordinate in the 1d effective action. This ambiguity is removed by applying the 
zero-energy constraint. 
The explicit form of the 1st order equations (\ref{eq:1st}) is highly dependent on the radial coordinate choice.
In our paper we will fix this ``gauge freedom'' by requiring that even in the deformed solution the
$G_{\tau\tau}$ and $G_{55}$ entries of the metric 
will remain equal exactly as in the supersymmetric case.
We will see that with this choice the set of the equations (\ref{eq:1st}) possesses an analytic solution.
On the contrary the radial coordinate ($\tau_\star$) of \cite{Borokhov:2002fm} is related to our coordinate ($\tau$) 
via $d \tau_\star = e^{4 \bar{p} - 4 \bar{q}} d \tau$. Note, however, that 
since both $\bar{p}(\tau)$ and
$\bar{q}(\tau)$ are expected to vanish at $\tau \to 0$ and $\tau \to \infty$, the 
deep UV and IR expansions of the fields have to be the same in terms of  $\tau$ and $\tau_\star$.

Let us first consider the equations of motion for $\xi_i$'s \footnote{We will set $g_s=1$
throughout this section}. Throughout this paper we will be
interested in a solution satisfying:
$ \xi_{Y}=\xi_{p}=\xi_{q}=0$.
Under this assumption we get:
\begin{eqnarray}   \label{eq:}
&\dot{\xi}_{y} = \xi_y \cosh y_0 + 2  e^{2 y_0} (2P -\tilde{F}_0) \xi_{\tilde{f}}-  &
\nonumber \\
&  - 2  e^{-2 y_0} \tilde{F}_0 \xi_{\tilde{k}},
\qquad
\dot{\xi}_{\tilde{f}+\tilde{k}} = 0, &
\nonumber \\
& \dot{\xi}_{\tilde{F}} = 
       - \cosh(2y_0) \xi_{\tilde{f}-\tilde{k}} - \sinh(2y_0) \xi_{\tilde{f}+\tilde{k}}, &
\nonumber \\
& \dot{\xi}_{\Phi} =   \left( e^{2y_0} (2P-\tilde{F}_0) \xi_{\tilde{f}} 
                                + e^{-2y_0} \tilde{F}_0 \xi_{\tilde{k}}   \right)- &
\nonumber \\
&    -  \frac{\tilde{k}_{0}-\tilde{f}_{0}}{2} \xi_{\tilde{F}},
\qquad
\dot{\xi}_{\tilde{f}-\tilde{k}} = - \xi_{\tilde{F}},   & 
\end{eqnarray} 
where $\xi_{\tilde{f} \pm \tilde{k}} = \xi_{\tilde{f}} \pm \xi_{\tilde{k}}$. 
We have $\xi_{\tilde{f}+\tilde{k}}=X$ for
constant $X$ and from the equations for 
$\xi_{\tilde{f}-\tilde{k}}$ and $\xi_{\tilde{F}}$ we obtain
a 2nd order differential order equation for $\xi_{\tilde{f}-\tilde{k}}$.
This equation has a two dimensional space of solutions. However, solving for $\xi_y$,
plugging the result into the zero-energy constraint $\xi_i \dot{\phi}_0^i =0$ and requiring also regularity at
$\tau \to 0$
we pick up a unique simple solution $
\xi_{\tilde{f}-\tilde{k}}(\tau)=X \cosh \tau$ and therefore:
\begin{eqnarray} 
&
\xi_{\tilde{F}}=- X \sinh \tau, \quad
\dot{\xi}_{\Phi} =0, 
&
\nonumber \\
&
\xi_{y}= 2 P X (\tau \cosh \tau - \sinh \tau), 
&
\end{eqnarray}
Having determined the explicit form of $\xi_i$'s we can consider the equations for
the fields $\bar{\phi}^i$'s.
For $\bar{y}$ we get:
\begin{equation}
\dot{\bar{y}} + \cosh (y_0) \bar{y} = \frac{2}{3} e^{4q_0 -4p_0-4Y_0} \xi_y.
\end{equation}
Using the result for $\xi_{y}$ and substituting the expressions for $q_0(\tau)$,
$p_0(\tau)$ and $Y_0(\tau)$ we may solve for $\bar{y}(\tau)$
fixing an integration constant by requiring regularity at $\tau \to 0$
(see \cite{Kuperstein:2003yt}).
In this review we will need an asymptotic behavior of $\bar{y}(\tau)$ at $\tau \to \infty$:
\begin{equation}     \label{eq:yUV}
\bar{y} \approx \mu \left( \tau - \frac{5}{2} \right) e^{-\tau/3} + V e^{-\tau} + \ldots,
\end{equation}
where $\mu$ is a deformation parameter proportional to $X$ and $V$ is a
numerical constant proportional to $\mu$.
Note that $\mu$ is a dimensionless parameter.
Using the result for $\bar{y}(\tau)$ and the fact that $\xi_{p}=0$ we may find the solution for 
$\bar{p}(\tau)$.
We refer the reader to the original paper \cite{Kuperstein:2003yt} for a full analytic solution 
for $p(\tau)$ and other fields (in particular it appears that $\bar{\Phi}=0$).
Here we will only review  the derivation of the results for the 3-form fields. 
Using the expressions for $\xi_{\tilde{f} \pm \tilde{k}}$
and $\xi_{\tilde{F}}$, passing from  $\tilde{f}$,  $\tilde{k}$
and $\tilde{F}$ to $f$, $k$ and $F$ we obtain and recalling that $\bar{\Phi}=0$:
\begin{eqnarray} \label{eq:fkF2}
&\dot{\bar{f}\,\,} + e^{2y_0} \bar{F} - 2 \dot{f}_0 \bar{y} = - \frac{2X}{2P} h_0 (\cosh \tau -1) &
\nonumber \\
&\dot{\bar{k}} - e^{-2y_0} \bar{F} + 2 \dot{k}_0 \bar{y} = \frac{2X}{2P} h_0 (\cosh \tau +1)&
\nonumber \\
&\dot{\bar{F}\,} - \half( \bar{k} -\bar{f} ) = - \frac{2X}{2P} h_0 \sinh \tau.&
\end{eqnarray}
Before discussing the explicit solution of this system it is worth to re-derive these equations
using the 2nd order type IIB equations of motion. In the most general ansatz preserving the global symmetry  
the 5-form $\tilde{F}_5$ is given by
\begin{equation}            
\tilde{F}_5 = \frac{1}{g_s} (1 + \star_{10}) \mathbf{d}\varphi \wedge \mathbf{d}x_0 \wedge \ldots
                               \wedge \mathbf{d}x_3,
\end{equation}
where $\varphi=\varphi(\tau)$. Supersymmetry requires
$\varphi=h^{-1}$ (see \cite{Grana:2000jj} and \cite{Gubser:2000vg}) , but it does not necessarily 
hold in a non-supersymmetric case. In what follows we will demonstrate how
assuming that $\dot{\Phi}=0$ and $\varphi=h^{-1}$ one may reproduce (\ref{eq:fkF2})
from the usual 2nd order 3-forms equations of
motion.  Indeed, under these assumptions the type IIB 3-forms equations
reduce to (\ref{eq:F3H3}). Let us expand (\ref{eq:F3H3}) around the supersymmetric KS solution. 
Note that the expansion includes also $\star_6$ due to the deformation of the 6d space.
We will denote the modified Hodge star operation by $\star_6 = \star_6^{(0)} + \bar{\star}_6$, where 
$\star_6^{(0)}$ corresponds to the supersymmetric configuration.
After some algebra the linearized RR 3-form equation reduces to:
\begin{eqnarray}   \label{eq:Z1}
&\mathbf{d}Z_3 =0,
\quad
\textrm{where} &
\nonumber \\
& Z_3 = \frac{1}{g_s} \bar{H}_3 + \star_6^{(0)} \bar{F}_3 +\bar{\star}_6 F^{(0)}_3.&
\end{eqnarray}
where $F_3^{(0)}$ is the RR 3-form in the KS background.
Similarly, from the NSNS 3-form equation we have:
\begin{equation}   \label{eq:Z2}
\mathbf{d} \star_6 Z_3 =0.
\end{equation}
Comparing this with (\ref{eq:fkF2}) we see that the r.h.s. of  (\ref{eq:fkF2}) is
exactly the components of the closed (and co-closed) form $Z_3$. 
Notice that having $Z_3 \neq 0$ necessary means that the complex form $G_{3}= F_{3}+ \frac{i}{g_s} H_{3}$
is not imaginary self dual and therefore the supersymmetry is broken \cite{Grana:2000jj}, \cite{Gubser:2000vg}.
The most general solution of (\ref{eq:Z1})
and (\ref{eq:Z2})
has 3 integration constants and it appears in
\cite{Kuperstein:2003yt}. In particular, it turns out that the 3-form on the r.h.s. of (\ref{eq:fkF2}) 
corresponds to the divergent $(0,3)$-form we have mentioned in the discussion following (\ref{eq:F3andH3}).
 Remarkably, this is the only solution for $Z_3$, which is consistent with $\dot{\Phi}=0$.
To find the solution for $\bar{F}(\tau)$,  $\bar{f}(\tau)$ and $\bar{k}(\tau)$
note that the homogeneous part of (\ref{eq:fkF2}) reduces to an
equation of the form $d Z_3 =d \star_6 Z_3 =0$ and as we have  already
mentioned the related 3-parameter solution appears in
\cite{Kuperstein:2003yt}.
Using this solution we may easily find the solution of the
three inhomogeneous equations (see \cite{Kuperstein:2003yt}). 
In the UV we have:
\begin{eqnarray}
&\bar{F}(\tau)  \approx  \mu \left( \frac{3}{4} \tau - 3 \right) e^{-\tau/3} + 
                               \left( \frac{3}{2} V + V^\prime \right) e^{-\tau} &
\nonumber \\
&\bar{f}(\tau) \approx  - \frac{27}{16} \mu e^{-\tau/3} + 
                               \left( \frac{V}{2} + V^\prime \right) e^{-\tau} + \ldots&
\nonumber \\
&\bar{k}(\tau) \approx   \frac{27}{16} \mu e^{-\tau/3} - 
                               \left( \frac{V}{2} + V^\prime \right) e^{-\tau} + \ldots,&
\nonumber 
\end{eqnarray}
where $V^\prime$ is a constant proportional to $\mu$.

Let us summarize. 
The deformation is controlled by the single parameter $\mu$ and all the fields have a regular behavior
in the UV and in the IR. There are two non-normalizable modes.
The first one is $y(\tau)$ and it is related to the deformation of the 6d metric. The second one 
is $\xi_2$ and it is associated with the 3-forms. In the UV we have:
\begin{equation}
\xi_2 \sim -F+\frac{k-f}{2} \approx -\frac{3}{4} \mu \left( \tau - \frac{25}{4} \right) e^{-\tau/3}.  
\end{equation}
Both $y(\tau)$ and  $\xi_2$ have dimension $\Delta=3$ which matches perfectly with the
asymptotic behavior of the fields. In the dual gauge theory these operators are dual to the gaugino 
bilinears. The deformation also involves other fields like $s=f+k$ with a normalizable 
behavior at $\tau \to \infty$. For example, $s \approx e^{-4 \tau/3}$ as expected for
an operator with $\Delta=4$.

\section{Vacuum energy}
\label{section:Ve}

To calculate the vacuum energy of the deformed non-supersymmetric theory 
we will use the standard AdS/CFT technique \cite{Aharony:1999ti}.
The supergravity dual of the gauge theory Hamiltonian is a $G_{00}$
component of the 10d metric. The vacuum energy, therefore, can be found
by variation of the type IIB SUGRA action  with respect to $G_{00}$.
This variation vanishes on-shell, except a boundary term. Looking at the supergravity
action, it is clear that the only such a boundary term will appear from the curvature part
of the action. Since the vacuum energy does not depend on the world-volume coordinates
we might consider the metric variation in the form
$G_{00} \to q G_{00}$.  
Here we only quota the final result for the vacuum energy (see \cite{Kuperstein:2003yt} for
the derivation):
\begin{equation}
E \sim \lim_{\tau \to \infty} \left( e^{n(\tau)} \partial_\tau \ln  h(\tau) \right),
\end{equation}
where
$\bar{n} = -4 \bar{q} + 4 \bar{p} + 4 \bar{Y}$.
The divergent result we have found is expected to 
be canceled out  when we compare the vacuum energies of
our solution and of the KS background, which we take as a reference. Using that
$h \to h_0 + \bar{h}$  and $n \to n_0 + \bar{n}$  
we get:
\begin{equation} \label{eq:Dexxx}
\Delta E \sim \left[ e^{n_0} \left(   \partial_\tau \left(\frac{\bar{h}}{h_0}\right) 
                                    +\bar{n}  \partial_\tau \left(\ln h_0\right)  \right) 
                \right]_{\tau \to \infty}, 
\end{equation} 
so that $\Delta E \sim \mu$.
Here we used the asymptotic solutions for the fields at $\tau \to
\infty$ from the previous section.
In (\ref{eq:Dexxx}) the term $e^{n_0(\tau)}$ diverges at $\tau \to \infty$ as $e^{4\tau/3}$. This is
suppressed by the $e^{-4\tau/3}$ term in the large $\tau$ expansion of
the fields appearing in the parenthesis  which multiply  the
$e^{n_0(\tau)}$ term. 
Furthermore, the term linear at $\tau$ cancels and we end up with a
constant proportional to $\mu$.

\section{Dual gauge theory}
\label{section:Dgt}

As was announced in the introduction the deformation of the supergravity
background corresponds in the gauge theory to 
an  insertion of the soft supersymmetry breaking gaugino mass terms.
The most general gaugino bilinear term has the form of
$\mu_+ {\cal O}_+ +\mu_-{\cal O}_- + c.c$ where ${\cal O}_\pm\sim
Tr[W_{(1)}^2 \pm W_{(2)}^2]$ and $W_{(i)}, \ i=1,2 $ relate  to the
$SU(N+M)$ and $SU(N)$ gauge groups respectively. Namely, the general
deformation is characterized by two complex masses. Our
non-supersymmetric deformation of the KS solution derived above
is a special case that depends on only one real parameter $\mu$.
Since the supergravity identification of the operators ${\cal O}_\pm$
is known up to some constants of proportionality we can not determine
the precise form of the soft symmetry breaking term. 

In the non-deformed supersymmetric 
theory the $U(1)_R$ symmetry is broken \cite{Klebanov:2002gr}, \cite{Herzog:2002ih}
first by instantons to $\mathbf{Z}_{2M}$ and then further
spontaneously broken down to
$\mathbf{Z}_2$
by a VEV of the gaugino bilinear. Let us discuss first the latter
breaking. 
We have already seen that on the SUGRA side this fact is manifest from 
the UV behavior of the complex 3-form $G_{3}= F_{3}+ \frac{i}{g_s} H_{3}$.
The sub-leading term in the expansion of $G_{3}$ preserves only the $\mathbf{Z}_2$
part of the $U(1)_R$ symmetry and it vanishes at infinity like $e^{-\tau}$ matching the 
expectation from the scalar operator $\textrm{Tr}(\lambda\lambda)$ of
dimension 3 with a non-zero VEV
\cite{Loewy:2001pq}.
Plugging the non-supersymmetric solution into $G_{3}$ we find that the leading term breaking
the  $U(1)_R$ symmetry behaves like $\Delta G _3 = g(\tau) e^{-\tau/3}$, where $g(\tau)$
is some polynomial in $\tau$. This is exactly what one would predict for an operator with
$\Delta=3$ and a non-trivial mass. The second combination of the gaugino bilinears
is encoded in the 6d part of the metric. For the 6d metric in (\ref{eq:ansatz}) to preserve
the $U(1)_R$ one has to set $y=0$. In the supersymmetric deformed conifold metric 
$y(\tau)= -2 e^{-\tau} + \ldots$ similarly to the behavior of the 3-form.
In the non-supersymmetric solution $y(\tau)$ goes like $e^{-\tau/3}$ elucidating again
that the gaugino combination gets a mass term. Notice also that the non-zero
VEVs of the gaugino bilinears are modified by the SUSY
breaking deformation. This is evident, for example, from the
$Ve^{-\tau}$ term in  the UV
expansion of $\bar{y}(\tau)$ in (\ref{eq:yUV}). Clearly, for $V \neq 0$
we have a correction to the VEV in the supersymmetric theory which was
encoded in the expansion of $y_0(\tau)$. Similar $e^{-\tau}$ term appears also in the
expansion of $\xi_2(\tau)$ and therefore the VEV of the second combination
of the gauginos gets modified too.

The  spontaneous  breaking of the $\mathbf{Z}_{2M}$ discrete group   down to the $\mathbf{Z}_2$ subgroup by
gaugino condensation results in an  
 $\mathbf{M}$-fold degenerate   vacua.
 This degeneracy is generally lifted by soft breaking mass terms in the action.
For small enough masses one can treat the supersymmetry breaking as a perturbation
yielding (for a single gauge group)  the well-known result \cite{Masiero:1985ss}
that the difference in energy between a non-supersymmetric
solution and its supersymmetric reference is given by
$\Delta E \sim \textrm{Re} (\mu C)$,
where $\mu$ and $C$ are the mass and the gaugino condensate respectively. 
For the gauge theory dual of the  deformed KS solution the vacuum
energy will in general be proportional to $\textrm{Re} (a_+\mu_+ C_+
+a_-\mu_-C_- )$
where $C_\pm$ are  the expectation values of ${\cal O}_\pm$ and
$a_\pm$ are some proportionality constants. In the special deformation
we are discussing in this paper this reduces to $\mu \textrm{Re} (a_+ C_+
+a_-C_- )$.
 In the previous section we have derived a result using the SUGRA dual
 of the gauge theory
which has this structure.
For the softly broken MN background similar calculations were
performed by \cite{Evans:2002mc}. In their case the explicit linear
dependence on the condensate was demonstrated.

One of the properties of the supersymmetric gauge theory is 
the space-time independence of the correlation function of two gaugino bilinears.
This appears from the supergravity dual description  as follows \cite{Loewy:2001pq}.
Consider a perturbation of the complex 2-form:
\begin{equation} 
C_2 \to C_2 + y \omega_2, \quad
G_3 \to G_3 + y \omega_3 + \mathbf{d} y \wedge \omega_2,
\end{equation}
where $\omega_{2,3}$ are given by (\ref{eq:DG})
and  $y(x,\tau)$ has non-vanishing boundary values.
Plugging this forms into the relevant part of the type IIB action
and integration over $\tau$ will \emph{not} lead to a kinematic term $dy(x_1)dy(x_2)$
and therefore the corresponding correlation function will be space-time independent.
This derivation is only schematic since there is a mixing between the 3-form modes and
the modes coming from metric as we have seen in Section \ref{section:NseoKS}.
Notice, however that this simplified calculation will yield the kinetic term for
the deformed non-supersymmetric background, since 
the complex 3-form is not imaginary self dual in this case. Thus in the non-supersymmetric 
theory the correlation function will be time-space dependent as one would expect.

\section{The plane wave limit} 
\label{section:Tpwl}

In this section we will construct a Penrose limit of the non-supersymmetric background. 
Following \cite{Gimon:2002nr} we will expand the metric around a null geodesic that
goes along an equator of the $S^3$  at $\tau = 0$.
The parameter $\varepsilon$ appearing in the 6d metric of the deformed conifold
and the gauge group parameter $\mathbf{M}$ are both taken to infinity in the PL limit,
while keeping finite the mass of the glue-ball:
$M_{gb} \sim \frac{\varepsilon^{2/3}}{g_s \mathbf{M} \alpha^\prime}$.
The final result \cite{Kuperstein:2003yt} is:
\begin{eqnarray}
& ds^2 = - 4 dx_- dx_+  + dx_i^2 + dz^2 + du d\bar{u} + &
\nonumber \\
&    + dv d\bar{v} - m_0^2 \Big[ v \bar{v} +
                                  \left( \left( \frac{4a_1}{a_0} -\frac{4}{5}\right) 
                                                 - 8\frac{3^{2/3}}{135} \mu  \right) z^2 &
\nonumber \\
&                                  +  \left( \left( \frac{4a_1}{a_0} -\frac{3}{5}\right)
                                                 + 4\frac{3^{2/3}}{135} \mu \right) u \bar{u} \Big] dx_+^2,&
\end{eqnarray}
where
\begin{equation}          \label{eq:m0}
m_0^2 = \frac{3^{1/3} \varepsilon^{4/3}}{2 (g_s \mathbf{M} \alpha^\prime)^2 a_0} \left(1 + 2C_Y \right).
\end{equation}
Recall that $C_Y$ is a numerical constant proportional to $\mu$. As expected for $\mu=0$ we recover the result of
the supersymmetric case \cite{Gimon:2002nr}.  
We see that all the world-sheet masses ($m_v$,$m_z$ and $m_u$) depend on
the supersymmetry breaking parameter.
Under the Penrose limit the 3-forms read:
\begin{eqnarray}
\left( F_3\right)_{+ v \bar{v}} = 
\left(\frac{1}{3} + 4 \gamma \right)^{-1}   \left( F_3\right)_{+ u \bar{u}} 
    = \frac{3 i m_0}{\sqrt{2} g_s} \sqrt{\frac{a_1}{a_0}} &&
\nonumber \\
\left(H_3 \right)_{+ u \bar{v}}=
\left(H_3 \right)_{+ v \bar{u}} = 
\frac{i m_0}{\sqrt{2}} \sqrt{\frac{a_1}{a_0}} (1-6\gamma). &&
\nonumber
\end{eqnarray}

\section{The plane wave string theory and  the Annulons} 
\label{section:TplwtatA}

The  string  theory  associated with the plane wave background
described in the previous section 
is quite  similar to that associated with the PL limit of the 
KS background.  The bosonic sector includes three massless fields that correspond to the 
spatial directions on the world-volume of the D3 branes. Their masslessness is 
attributed to the translational invariance of the original metric and
the fact that the null geodesic is at constant $\tau$. The rest five
coordinates  are massive. 
The only difference between the bosonic spectrum of the 
deformed model and that of \cite{Gimon:2002nr} is the shift of the masses
of the $z,v,\bar v,u,\bar u$ fields. 
The sum of the $mass^2$, however,
of the individual fields $\sum m^2 = 12m_0^2\frac{a_1}{a_0}$ still has  the same 
form as the sum in the supersymmetric case apart from the modification of $m_0$ (\ref {eq:m0}).
The modification of $m_0$  is also responsible for   the deviation of
the deformed string tension with from  the supersymmetric one since
the string tension 
$T_s \sim g_s \mathbf{M}   m_0^2$.
The fermionic spectrum takes the form ($k=1,\ldots,4$ and $l=1,2$):
\begin{eqnarray}
&& \omega_n^k  \approx \sqrt{n^2 + \hat{m}_B^2 \left( 1 + 18 \gamma  \right)},
\nonumber \\
&& \omega_n^l = \sqrt{n^2 + \frac{1}{4} \hat{m}_B^2} \pm\frac{1}{2}\hat{m}_B ,
\quad \textrm{where}
\nonumber \\
&&\hat{m}_B =   \sqrt{2} p^+ \alpha^\prime m_0 \left( \frac{a_1}{a_0}\right)^{1/2} (1-6\gamma).
\nonumber
\end{eqnarray}
Comparing the bosonic and fermionic masses we observe that like in  the undeformed KS model 
there is no linearly realized world-sheet supersymmetry and the hence
there is a 
non-vanishing  zero point energy.
However,  up to deviations linear in $\mu$ the sum
of the square of the frequencies  of the bosonic and fermionic modes match. 
Since this property follows  in fact from the relation between $R_{++}$ and $\left( G_3 \right)_{+ij}  \left( \bar{G}_3 \right)_+^{\,\,ij}$  
it should be  a universal property  of any plane wave background.  

Surprisingly we find that the fermionic spectrum admits  two fermionic zero modes $\omega_0^{l=1,2}$
exactly like in the supersymmetric case.
The fermionic zero modes in the spectrum of the latter case  
were predicted \cite{Gimon:2002nr} upon observing  that 
the Hamiltonian still commutes with the four supercharges that correspond to the four dimensional
${\cal N}=1$ supersymmetric gauge theory. 
This implies that four supersymmetries out of the sixteen
supersymmetries of plane wave solution
commute with the Hamiltonian giving rise to the four zero-frequency modes
and a four dimensional Hilbert sub-space of (two bosonic and two fermionic) degenerate states.
One might have expected that in the PL of the deformed  theory
the fermionic zero modes will be lifted by an amount proportional to the supersymmetry breaking parameter.
Our results, however, contradict this expectation.
In the dual field theory this implies that even though the full theory is non-supersymmetric,
the sector of states with large $J$ charge  admits supersymmetry. As will be
discussed below  these states
are characterized by their large mass which is proportional to $J$. Presumably, in this limit of states of large mass the 
impact of the small  gaugino mass deformations is washed away. For instance one can estimate that the ratio of the  boson 
fermion mass difference to the mass of the annulon scales like $ \frac{\mu}{J}$ and since $\mu$ has to be small and $J\rightarrow \infty$ this ratio is negligible.

Note that the fermionic zero modes are in accordance with the    the criteria   presented in
\cite{Apreda:2003gs}. However, the metric and the 3-form given
in \cite{Apreda:2003gs} do not coincide   with our results, because of the factor of $C_Y$ in the
expression for $m_0^2$.

Since apart from the modifications  of the fermionic and bosonic
frequencies the string Hamiltonian we find
has the same structure as the one found for the KS case,  the analysis of the corresponding
gauge theory states also follows that of  \cite{Gimon:2002nr}. 
We will not repeat here this analysis, but rather just summarize its outcome:
\begin{itemize}
\item
The ground state of the string corresponds to the {\it Annulon}. 
This hadron which carries a large $J$ charge 
is also very massive since  its mass is given by 
\begin{eqnarray}
M_{annulon} = m_0 J 
 \end{eqnarray} 
Obviously, the only difference between the annulon of the deformed theory in comparison with the 
supersymmetric one is the modification of $m_0$.
\item
The annulon can be described as  a ring 
composed of $J$ constituents each having a mass ( in the mean field of all the others)  of $m_0$.
\item
The annulon which is a space-time scalar has a fermionic superpartner of the same mass. The same holds for the rest
of the bosonic states.
\item
The string Hamiltonian has a term $ \frac{P_i^2}{2m_0J}$ that describes a non-relativistic motion of the annulons.
\item
The annulons admit stringy ripples. The spacing between these
excitations are proportional 
to $\frac{T_s} {M_{annulon}}$.
\item 
The string Hamiltonian describes also excitations that correspond to
the addition of small number of different constituents on top of the J basic ones.
\end{itemize}

\section{Acknowledgments}
The author thanks G. 't Hooft for the opportunity to speak in the new
talent sessions at the 41st International School on Subnuclear Physics in Erice, Italy.

\bibliographystyle{JHEP}
\bibliography{Erice}

\end{document}